\documentclass[12pt]{article}
\input epsf.tex
\usepackage{color}

\renewcommand{\thefootnote}{\fnsymbol{footnote}}
\newcommand{\beq}{\begin{equation}}
\newcommand{\eeq}{\end{equation}}
\newcommand{\beqa}{\begin{eqnarray}}
\newcommand{\eeqa}{\end{eqnarray}}
\newcommand{\beqar}{\begin{eqnarray*}}
\newcommand{\eeqar}{\end{eqnarray*}}

\newcommand{\eps}{\epsilon}
\newcommand{\ga}{\gamma}

\newcommand{\inn}{\!\cdot\!}

\newcommand{\eg}{{\it e.g.,}\ }
\newcommand{\ie}{{\it i.e.,}\ }
\newcommand{\labell}[1]{\label{#1}} 
\newcommand{\reef}[1]{(\ref{#1})}
\newcommand\prt{\partial}
\newcommand\veps{\varepsilon}

\newcommand\cR{{\cal R}}
\newcommand\cA{{\cal A}}
\newcommand\cM{{\cal M}}
\newcommand\cN{{\cal N}}

\newcommand\cJ{{\cal J}}

\newcommand\cL{{\cal L}}
\newcommand\cG{{\cal G}}
\newcommand\cE{{\cal E}}
\newcommand\cI{{\cal I}}

\newcommand\cP{{\cal P}}

\newcommand\bz{\bar{z}}

\newcommand\Tr{{\rm Tr}}

\begin{document}

\vspace*{1cm}

\begin{center}
{\bf \Large On Chern-Simons Couplings at order $O(\alpha'^2)$}

\vspace*{1cm}

{Komeil Babaei Velni$^{1}$\footnote[1]{babaeivelni@guilan.ac.ir} and Ali Jalali$^{2}$}\footnote[2]{ali.jalali@stumail.um.ac.ir}\\
\vspace*{1cm}
$^{1}${Department of Physics, University of Guilan,\\ P.O. Box 41335-1914, Rasht, Iran}
\\
\vspace{0.5cm}
$^{2}${Department of Physics, Ferdowsi University of Mashhad,\\ P.O. Box 1436, Mashhad, Iran}
\\
\vspace{2cm}

\end{center}

\begin{abstract}
\baselineskip=18pt

Using the explicit string scattering calculation and the linear T-dual ward identity, we evaluate the string theory disc amplitude of one Ramond-Ramond field $C^{(p+1)}$
and two Neveu-Schwarz B-fields in the presence of a single $D_p$-brane in type $IIB$ string theory. From this amplitude we extract the $O(\alpha'^2)$ (or equivalently four-derivative) part of the $D_p$-brane couplings involving these fields.

\end{abstract}

\vfill
\setcounter{page}{0}
\setcounter{footnote}{0}
\newpage
\renewcommand{\thefootnote}{\arabic{footnote}}
\section{Introduction} \label{intro}
 Achieving a detailed  understanding of string theory vacua  plays a very important role in the relationship between this theory with real-world physics. One of the most ingredients of vacuum is $D$-brane (localized objects on which strings can end).  
$D$-branes are non-perturbative excitations in the theory which are localized to a sub-manifold of the ten-dimensional  space-time. We will be considering the stable D-branes that carry the $RR$ charges. There are many degrees of freedom localized on the $D$-brane's worldvolume.

Most of $D$-brane's  researches being done focuses on  their low-energy dynamics\color{red}, \color{black} and so the effects of higher derivative
corrections to the string theory action are often ignored. However, it has been shown that the corrections of high order, has a very important role in establishing the consistency, or inconsistency, of some solutions. The investigation of these solutions demonstrate that the  higher derivative
corrections to the D-brane actions can also have a major role in determining the consistency of string  compactifications. Regardless of these corrections, understanding the proper structure of the string vacua is not possible \cite{Becker:1996np,Becker:2010ij,Becker:2011ad}.

BPS $D_p$-branes in type $II$ string theories ($p$ even(odd) in $IIA$($IIB$) theory) are stable solitons which break half of the space-time supersymmetries and their dynamics are properly described in field theory by Dirac-Born-Infeld (DBI) action and Chern-Simons action (CS)  \cite{Leigh:1989jq,Bachas:1995kx}.

The DBI action  describes the dynamics of the brane in the presence of $NSNS$ background fields. The CS part describes the coupling of $D$-branes to the $RR$ potential. For constant fields it is given by \cite{Polchinski:1995mt,Douglas:1995bn}
\beqa
S_{CS}&=&T_{p}\int_{M^{p+1}}e^{B}C\labell{CS2}
\eeqa
where  $M^{p+1}$ represents the world volume of the D$_p$-brane, $C$ is the sum over all  RR potential forms, \ie $C=\sum_{n=0}^8C^{(n)}$,  and the multiplication rule is  wedge product. The abelian gauge field can be added to the action as $B\rightarrow B+2\pi\alpha'f$. Curvature correction to this action has been found in \cite{Green:1996dd,Cheung:1997az,Minasian:1997mm}  by requiring that the chiral anomaly on the world volume of intersecting $D$-branes cancels  the anomalous variation of the CS action.

Additional linear couplings to the $NSNS$ fields were found by studying the S-matrix element of one $RR$ and one $NSNS$ vertex operator  at order $O(\alpha'^2)$ \cite{Garousi:1996ad}. These couplings take the following form  \cite{Garousi:2010ki}:
\beqa
 &&\pi^2\alpha'^2T_p\int d^{p+1}x\,\eps^{a_0\cdots a_p}\bigg(\frac{1}{2!(p-1)!}[{ F}^{(p)}_{ia_2\cdots a_p,a}H_{a_0a_1}{}^{a,i}-{ F}^{(p)}_{aa_2\cdots a_p,i}H_{a_0a_1}{}^{i,a}]\nonumber\\
&&\qquad\qquad\qquad\qquad\quad+\frac{2}{p!}[\frac{1}{2!}{ F}^{(p+2)}_{ia_1\cdots a_pj,a}\cR^a{}_{a_0}{}^{ij}-\frac{1}{p+1}{ F}^{(p+2)}_{a_0\cdots a_pj,i}(\hat{\cR}^{ij}-\phi\,^{,ij})]\nonumber\\
&&\qquad\qquad\qquad\qquad\quad-\frac{1}{3!(p+1)!}{ F}^{(p+4)}_{ia_0\cdots a_pjk,a}H^{ijk,a}\bigg)\labell{LTdual}
\eeqa

where the commas are used to denote partial differentiation, $\cR$ is the linearized curvature tensor  and $F^{(n)}=dC^{(n-1)}$. The above couplings are invariant under the linear $T$-dual ward identity corresponding to $NSNS$ gauge transformations\footnote{Extension of the  couplings in \reef{LTdual} to one RR, one NSNS and one NS open string couplings have been studied in \cite{jalali}.}.

These quadratic  couplings can be extended to higher order terms by making them  covariant under the coordinate transformations and invariant under the following $RR$ gauge transformations:
\beqa
\delta C=d\Lambda+H\wedge \Lambda \labell{RR}
\eeqa
  where $H$ is the field strength of $B$, \ie $H=dB$, and $\Lambda=\sum_{n=0}^7\Lambda^{(n)}$. That is, the partial derivatives  in these couplings should be replaced by the covariant derivatives, the closed string tensors should be extended to the pull-back of the bulk fields onto the world-volume of $D$-brane, the linear $RR$ gauge field strength  should be extended to the nonlinear field strength $\tilde{F}^{(n)}=dC^{(n-1)}+H\wedge C^{(n-3)}$, and the linear curvature tensor should be extended to the nonlinear curvature tensor. These couplings for ${\cal C}^{(p+1)}$ are\cite{Garousi:2010rn}:
 \beqa
&&\frac{T_{p}}{(p-1)!}\int d^{p+1}x\epsilon^{a_0\cdots a_{p-2}bd}{\cal C}^{(p+1)}_{a_0\cdots a_{p-2}}{}^{ij}\bigg[\frac{1}{2}H_j{}^{fe}{}_{,b}H_{ied,f}+\frac{1}{4}\cR_{bd}{}^{ef}\cR_{ijfe}+\cR^e{}_{jb}{}^k\cR_{eidk}\nonumber\\
&&\qquad\qquad\qquad\qquad\qquad\qquad\qquad-\frac{1}{2}H_j{}^{lk}{}_{,b}H_{ikd,l}-\frac{1}{4}\cR_{bd}{}^{kl}\cR_{ijlk}-\cR^k{}_{jb}{}^e\cR_{kide}\bigg]\labell{new25}
\eeqa 

In the present paper, we are interested in generalizing the CS action to take into account some other couplings of $C^{(p+1)}\wedge H\wedge H$.  To compute the terms of interest, we must evaluate scattering amplitudes in which various closed string fields interact with a D-brane. We will restrict ourselves to  tree-level computations, so the relevant amplitudes are given by insertions of multiple closed string vertex operators on a world-sheet with the topology of a disc. From these S-matrix elements, we conjecture an extension for the CS action which includes these  couplings as well. In fact, we will present some four-derivative corrections involving a $RR$ potential of degree $p+1$, and two $NSNS$ fields.

The outline of this paper is as follows. In section 2 we start by  writing the S-matrix elements of one $RR$ and two $NSNS$ vertex operators. From this approach, we find the S-matrix elements for the $RR$  potential with degree $(p+1)$ that carry four and three transvers indices. we check the consistency of these elements with the results in \cite{Komeil:2013prd} where the  relevant amplitudes found by using of T-dual ward identity. In section 3, by applying the explicit scattering calculation,
 we drive the amplitude for $RR$ $(p+1)$-form with one transverse index in terms of the $RR$ structures. Next, in section 4, we calculate the amplitude in low energy limit and find some new couplings.


\section{The approach for three closed string amplitude; one $RR$ and two $NSNS$}
 The explicit  calculations  of the scattering amplitude of one RR n-form in (-1/2, -3/2)-picture and two NSNS vertex operators in  (0, 0)-picture on the world volume of D$_p$-brane has been studied in \cite{Becker:2011ad, Garousi:2010bm, Garousi:2011bm, Komeil:2015}. We review this calculations and apply them to find the  interesting amplitudes. We work in the RNS  world-sheet formalism, with the closed string vertex operators being  constructed out of  bosons  $X^{\mu}(z), X^{\mu}(\bz)$ and fermions $\psi^{\mu}(z), \psi^{\mu}(\bz)$ as well as the picture ghosts $\phi(z)$ and $\phi(\bz)$. 
 The scattering amplitude is given by the following correlation function:
\beqa
\cA&\sim&<V_{RR}^{(-1/2,-3/2)}(\veps_1^{(n)},p_1)V_{NSNS}^{(0,0)}(\veps_2,p_2)V_{NSNS}^{(0,0)}(\veps_3,p_3)>\labell{amp2}
\eeqa
Using the doubling trick \cite{Garousi:1996ad} and standard  relevant vertex operators \cite{Billo:1998vr}, explicitly we have
\beqa
\cA&\sim&\frac{1}{2}(H_{1(n)}M_p)^{AB}(\veps_2\inn D)_{\mu_3\mu_4}(\veps_3\inn D)_{\mu_5\mu_6}\int d^2z_1d^2z_2d^2z_3\, (z_{1\bar 1})^{-3/4}\nonumber\\
&&(\sum_{i = 1 }^ {10}b_{i}+\sum_{i = 1 }^ {6}b'_{i}) ^{\mu_3\mu_4\mu_5\mu_6}_{AB}\delta^{p+1}(p_1^a+p_2^a+p_3^a)+(2\leftrightarrow 3)\labell{A}
\eeqa
 
 where the NSNS vertex operators include polarizations $\veps_i$  and momenta $p_i$ ,$i=2,3$. The RR vertex operator has momentum  $p_1$ and antisymmetric polarization $\veps_1$. The indices $A,B,\cdots$ are the Dirac spinor indices, and
\beqa
H_{1(n)}&=&\frac{1}{n!}\veps_{1\mu_1\cdots\mu_{n}}\gamma^{\mu_1}\cdots\gamma^{\mu_{n}}\nonumber\\
M_p&=&\frac{\pm 1}{(p+1)!} \eps_{a_0 \cdots a_p} \ga^{a_0} \cdots \ga^{a_p}\labell{H1}
\eeqa
where $\eps$ is the volume $(p+1)$-form of the $D_p$-brane. Here the matrix $D_{\mu\nu}$ is a diagonal matrix that agrees with $\eta_{\mu\nu}$ in directions along the
brane (Neumann boundary conditions) and with $-\eta_{\mu\nu}$  in directions normal to the brane (Dirichlet boundary conditions). In this notation, $D^{\mu i}=-\delta^{\mu i},  D^{\mu a}=\delta ^{\mu a}$ and thus $D_{\mu\nu}=V_{\mu\nu}-N_{\mu\nu}$. One has to use the standard world-sheet propagators to write amplitude \reef{A}.$b_i$'s are some correlation functions. These functions include the correlators of $X$'s, the correlators of two spin operators and the correlation function involving two spin operators and $\psi$'s. $b_i$'s have been illustrated in \cite{Garousi:2010bm, Garousi:2011bm}. $b'_i$'s are also some correlation functions that contribute to the amplitude of our interested cases in this paper
\begin{eqnarray}\labell{bi}
(b'_{11})^{\mu_3\mu_4\mu_5\mu_6}_{AB}&=&(ip_3)_{\beta_1}<:S_A:S_B:\psi^{\beta_1}\psi^{\mu_5}:>g_{11}^{\mu_3\mu_4\mu_6}\cr
(b'_{12})^{\mu_3\mu_4\mu_5\mu_6}_{AB}&=&(ip_3\inn D)_{\beta_1}<:S_A:S_B:\psi^{\beta_1}\psi^{\mu_6}:>g_{12}^{\mu_3\mu_4\mu_5}\nonumber\\
(b'_{13})^{\mu_3\mu_4\mu_5\mu_6}_{AB}&=&(ip_2\inn D)_{\beta_1}(ip_3)_{\beta_2}<:S_A:S_B:\psi^{\beta_1}\psi^{\mu_4}:\psi^{\beta_2}\psi^{\mu_5}:>g_{13}^{\mu_3\mu_6}\nonumber\\
(b'_{14})^{\mu_3\mu_4\mu_5\mu_6}_{AB}&=&(ip_3)_{\beta_1}(ip_3\inn D)_{\beta_2}<:S_A:S_B:\psi^{\beta_1}\psi^{\mu_5}:\psi^{\beta_2}\psi^{\mu_6}:>g_{14}^{\mu_3\mu_4}\nonumber\\
(b'_{15})^{\mu_3\mu_4\mu_5\mu_6}_{AB}&=&(ip_2)_{\beta_1}(ip_3)_{\beta_2}(ip_3\inn D)_{\beta_3}<:S_A:S_B:\psi^{\beta_1}\psi^{\mu_3}:\psi^{\beta_2}\psi^{\mu_5}:\psi^{\beta_3}\psi^{\mu_6}:>g_{15}^{\mu_4}\nonumber\\
(b'_{16})^{\mu_3\mu_4\mu_5\mu_6}_{AB}&=&(ip_2\inn D)_{\beta_1}(ip_3)_{\beta_2}(ip_3\inn D)_{\beta_3}<:S_A:S_B:\psi^{\beta_1}\psi^{\mu_4}:\psi^{\beta_2}\psi^{\mu_5}:\psi^{\beta_3}\psi^{\mu_6}:>g_{16}^{\mu_3}\nonumber
\end{eqnarray}
where $g$'s are  the correlators  of $X$'s which can easily be performed using the standard world-sheet propagators, and  
the correlator of $\psi$ can be calculated using the  Wick-like rule. Respecting the following relation between $b_i$ and $b'_i$ in \cite{Garousi:2010bm,Garousi:2011bm}, it could be found that the contribution of the above correlators appear as a overall factor of 2. 
\beqa
b'_{11}=b_2,\, \,\,\,\,\,\,b'_{12}=b_3,\,\,\,\, \,\,\,b'_{13}=b_6,\,\,\,\, \,\,\,b'_{14}=b_4,\,\,\,\, \,\,\,b'_{15}=b_8,\,\,\,\, \,\,\,b'_{16}=b_9.\nonumber
\eeqa
One can find that the amplitude involves the following trace that  appears by combining the gamma matrices in amplitude \reef{A} and the Wick-like rule gamma matrices  \cite{Garousi:2010bm}:
 \beqa
T(n,p,m)
& =&\frac{1}{n!(p+1)!}\veps_{1\nu_1\cdots \nu_{n}}\eps_{a_0\cdots a_p}A_{[\alpha_1\cdots \alpha_m]}\Tr(\gamma^{\nu_1}\cdots \gamma^{\nu_{n}}\gamma^{a_0}\cdots\gamma^{a_p}\gamma^{\alpha_1\cdots \alpha_m})\labell{trace}
 \eeqa
where $A_{[\alpha_1\cdots \alpha_m]}$ is an antisymmetric combination of the momenta and/or the polarizations of the $NSNS$ states and $m=0,2,4,6,8$. The scattering amplitude is non-zero only  for the following sequence \cite{Garousi:2010bm}:
\beqa
n=p-3,\, n=p-1,\, n=p+1,\, n=p+3,\, n=p+5 \labell{seq}
\eeqa
The explicit calculations in order to find the amplitudes corresponding to the first two terms of this sequence have been done in \cite{Becker:2011ad, Garousi:2010bm, Garousi:2011bm, Komeil:2015}. As we have illustrated in Table 1, we  separate the  known $RR$ corresponding amplitudes that has been  calculated explicitly in literatures. We are going to continue our efforts to find all $RR$ n-form amplitudes.  In this paper, we are interested in the case $n=p+1$. In this case 
we consider the $RR$ potentials with degree $n=p+1$ that carry four, three and two transverse indices. The  nonzero trace relation \reef{trace} for this case mentions in Table 1. 
\begin{table}[ht]
\caption{The calculated scattering amplitudes of 
one $RR$ and two $NSNS$ in explicit approach}
\centering
\begin{tabular} {c c c c c c c c c c c c c}
\hline \hline 
    &  transverse indices &\,\,\,  \,\,\, & nonzero trace  \,\,\,  &\,\,\, &\,\,\, &\,\,\, &          reference\\
\hline \hline 
n=p-3 \,\,\,\,\,\,\,\,\, &  2\,\,\, &\,\,\,\,\,\,\,\,\, & T(2, 5, 8)\,\,\,\,\,\,  &&&&  \cite{Garousi:2010bm}, \cite{Becker:2011ad}\\
\hline 
n=p-3 \,\,\,\,\,\,\,\,\, &  1\,\,\, &\,\,\,\,\,\,\,\,\, & T(1, 4, 6)\,\,\,\,\,\,  &&&&  \cite{Garousi:2010bm}, \cite{Becker:2011ad}\\
\hline 
n=p-3 \,\,\,\,\,\,\,\,\, &  0\,\,\, &\,\,\,\,\,\,\,\,\, & T(0, 3, 4)\,\,\,\,\,\,  &&&&  \cite{Garousi:2011bm}, \cite{Becker:2011ad}\\
\hline 
n=p-1 \,\,\,\,\,\,\,\,\, &  3\,\,\, &\,\,\,\,\,\,\,\,\, & T(3, 4, 8)\,\,\,\,\,\,  &&&&  \cite{Komeil:2015}\\
\hline 
n=p-1 \,\,\,\,\,\,\,\,\, &  2\,\,\, &\,\,\,\,\,\,\,\,\, & T(2, 3, 6)\,\,\,\,\,\,  &&&&  \cite{Komeil:2015}\\
\hline 
n=p-1 \,\,\,\,\,\,\,\,\, &  1\,\,\, &\,\,\,\,\,\,\,\,\, & T(1, 2, 4)\,\,\,\,\,\,  &&&&  \cite{Komeil:2015}\\
\hline 
n=p+1 \,\,\,\,\,\,\,\,\, &  4\,\,\, &\,\,\,\,\,\,\,\,\, & T(4, 3, 8)\,\,\,\,\,\,  &&&&   current paper\\
\hline 
n=p+1 \,\,\,\,\,\,\,\,\, &  3\,\,\, &\,\,\,\,\,\,\,\,\, & T(3, 2, 6)\,\,\,\,\,\,  &&&&  current paper\\
\hline 
n=p+1 \,\,\,\,\,\,\,\,\, &  2\,\,\, &\,\,\,\,\,\,\,\,\, & T(2, 1, 4)\,\,\,\,\,\,  &&&&  current paper
\end{tabular}
\end{table}

In the following, we will evaluate the amplitude corresponding to $n=p+1$ $RR$ potentials. Since the similar calculations for another cases of RR potentials have been performed  in the papers  mentioned above, thus  the summary  of results for  interesting cases  presented here. Let us begin with $\veps_1^{ijkl}$. 

\subsection{$n=4$ case}\labell{n4}

 From the relation $n=p+1$ for this RR potential $\veps_1^{ijkl}$, one gets $p=3$. The nonzero trace  \reef{trace} becomes 
\beqa
T(4,3,8)&=&32\frac{8!}{4!4!}\veps_1^{ijkl}\eps^{a_0 a_1 a_2 a_3}A_{[ijkla_0 a_1 a_2a_3]}\labell{T438}
\eeqa

where 32 is the trace of the $32\times 32$ identity matrix.
Since $m=8$,  only the $\psi$ correlator in $b_{10}$ have non-zero contribution to the amplitude \reef{amp2}. After mapping the results to disk with unit radius and set \cite{Becker:2010ij} $z_1=0$ to fix $SL(2,R)$ symmetry, it can be found the $X$ correlator in  $b_{10}$ is \cite{Garousi:2010bm, Garousi:2011bm}
  \beqa
|z_2| ^{2s}|z_3|
^{2t} \left(1-|z_2|^2 \right)^{p} \left(1-|z_3|^2\right)^{q}| z_2-z_3| ^{2u}| 1-z_2\bar
z_3| ^{2v}\equiv K\labell{K}
\eeqa

where we have used the following  definitions for the  Mandelstam  variables:
\beqa
 s=p_1\cdot p_2 \quad &&; \quad t=p_1\cdot p_3 \quad ; \quad u=p_2\cdot p_3 \cr
 p=p_2\cdot D \cdot p_2 \quad &&\quad ; \quad q=p_3\cdot D \cdot p_3 \quad ; \quad v=p_2\cdot D \cdot p_3 
\eeqa

Using the fact that the kinematic factor in \reef{T438} contract with $\veps_1^{ijkl}\eps^{a_0a_1a_2a_3}$, one observes  that there are 70 different terms, however, some of them are zero and finally two terms remain independently. Considering the $\psi$-correlator from Wick-like rule and the above $X$-correlator in \reef{A}, one finds that the scattering amplitude is nonzero only for the $NSNS$ polarization tensors with the same symmetry. This amplitude becomes\footnote{Our conventions set $\alpha'=2$ in the string theory amplitudes. Our index convention is that the Greek letters  $(\mu,\nu,\cdots)$ are  the indices of the space-time coordinates, the Latin letters $(a,d,c,\cdots)$ are the world-volume indices and the letters $(i,j,k,\cdots)$ are the normal bundle indices.}:
\beqa
{ \cA}&\sim& T_3\eps_{a_0\cdots a_3}p_2^{a_0}p_3^{a_1}\bigg(\frac{1}{4}(\veps_2^S\inn N\inn \cE\inn N\inn\veps_3^S)^{a_2a_3}+\Tr(\cE \inn N\inn \veps_2^A)(\veps_3^A)^{a_2a_3}\bigg)\nonumber\\
&&\times\cI_1\delta(p_1^a+p_2^a+p_3^a)+(2\leftrightarrow 3)\labell{A4ee}
\eeqa
where $T_3$ is the tension of  D$_3$-brane and  $\veps^A,\, \veps^S$ are the polarizations of the B-field and graviton respectively. We define $\cE^{nm}=(p_2\inn N\inn \veps_1\inn N\inn p_3)^{nm}$
 in which $n, m$ are the transvers indices.

The closed and open string channels appear in the integral $\cI_1$. The explicit form of this integral and it's low energy expansion has been found in \cite{Garousi:2010bm}. In future section, we will study all integrals that appear in the three closed string amplitudes. The above amplitude has been found by $T$-duality in \cite{Komeil:2012np}. 
\subsection{$n=3$ case}
In this case of $(p+1)$-form $RR$ potential $\veps_1^{ijk}$ that carries three transverse indices, we have $p=2$.  The $\psi$ correlators in $b_{8},b_{9},b_{10},b'_{15},b'_{16}$ have non-zero contribution to the amplitude \reef{amp2}. It could be found that this scattering amplitude is zero when the symmetry of $NSNS$ polarizations are different. So the amplitude is non-zero for two graviton or two B-field vertex operators. The trace relation \reef{trace} is non-zero only for $m=6$. The trace in this case becomes:
\beqa
T(3,2,6)&=&32\frac{6!}{3!3!}\veps_1^{ijk}\eps^{a_0 a_1 a_2}A_{[ijka_0 a_1 a_2]}\labell{T326}
\eeqa
Using the on-shell condition $\veps_2\cdot p_2=\veps_3\cdot p_3=0$ and conservation of momentum along the world volume, it  could be found  that the $X$ correlators corresponding to these $b_{i}$'s and $b'_{i}$'s. After calculating the sub-amplitudes ${\cal A}_i$ in \reef{A} corresponding to $b_{8},b_{9},b_{10},b'_{15},b'_{16}$, one can find that the polarization of $RR$ $(p+1)$-form with three transverse indices appears in amplitude in the form of following RR structures:
\beqa
&&Tr(\cE\inn N\inn \veps^A\inn N),    \quad(p\inn \veps^S \inn N\inn \cE\inn N\inn \veps^S)^{\mu},    \quad(\veps^S \inn N\inn \cE\inn N\inn \veps^S)^{\mu\nu},\quad(\cE\inn N\inn \veps^S)^{\mu}\nonumber\\
&&(p\inn \veps^A \inn N\inn \cE) ,  \qquad\quad(\cE\inn N\inn \veps^A\inn\veps^A)^{\mu},\quad\quad\quad(\cE\inn N\inn \veps^S\inn\veps^S)^{\mu}\nonumber
\eeqa
where $\mu,\nu$ are the world volume indices that contract with the volume $(p+1)$-form $\eps$. In above, we define $\cE^{nm}=(p_i\inn N\inn \veps_1)^{nm},\,\, i=2,3$ that appears in the first line and also $\cE^{n}=(p_i\inn N\inn \veps_1\inn N\inn p_j)^{n},\,\, i,j=2,3\,\,, i\neq j$ that appears in the other structures. $n, m$ are the transverse indices. $\veps^A,\, \veps^S$ are the polarizations of the B-field and graviton respectively. 

 We find the amplitude of two $B$-fields in terms of the above RR structures:
 \beqa
{\cal A}_{A} &\sim& \eps_{a_0a_1a_2}\bigg[Tr(\cE\inn N\inn \veps_2^A\inn N) \cM_1^{a_0a_1a_2}+Tr(\cE\inn N\inn \veps_3^A\inn N) \cM_2^{a_0a_1a_2}\nonumber\\
&&\quad\quad+(p_1\inn N\inn \veps_3^A \inn N\inn \cE) \cM_3^{a_0a_1a_2}+(p_2\inn N\inn \veps_3^A \inn N\inn \cE) \cM_4^{a_0a_1a_2}\nonumber\\
&&\quad\quad+(p_2\inn V\inn \veps_3^A \inn N\inn \cE) \cM_5^{a_0a_1a_2}+(p_3\inn V\inn \veps_2^A \inn N\inn \cE) \cM_6^{a_0a_1a_2}\nonumber\\
&&\quad\quad+(p_3\inn N\inn \veps_2^A \inn N\inn \cE) \cM_7^{a_0a_1a_2}+(\cE\inn N\inn \veps_2^A\inn V\inn \veps_3^A)^{a_2} \cM_8^{a_0a_1}\nonumber\\
&&\quad\quad+(\cE\inn N\inn \veps_2^A\inn N\inn \veps_3^A)^{a_2} \cM_9^{a_0a_1}\bigg]+(2\leftrightarrow 3)\labell{AlastA}
\eeqa
 where the  $\cE$ tensors (contraction of $NSNS$ momenta with the $RR$ polarization tensor) in the above S-matrix elements that appear in the bracket are $\cE^{nm}=(p_2\inn N\inn \veps_1)^{nm}$ and   $\cE^{n}=(p_2\inn N\inn \veps_1\inn N\inn p_3)^{n}$. Thus we have  $\cE^{nm}=(p_3\inn N\inn \veps_1)^{nm}$ and  $\cE^{n}=(p_3\inn N\inn \veps_1\inn N\inn p_2)^{n}$ in the $(2\leftrightarrow 3)$. The subscript $A$ refers to the amplitude that involves two antisymmetric $NSNS$ polarization tensors.
 
  T-duality could find the first line of the above antisymmetric amplitude where only one index of the RR potential contracts with the momentum corresponding to NSNS polarizations and the other indices of the RR potential contract with the NSNS polarization.   
 
 Since there is no conservation of  transverse momentums in \reef{A}, the terms including contraction of $p_1^i$ with each of the polarizations, \ie  $p_1\inn N\inn\veps_3,\, p_1\inn N\inn\veps_2$, as well as  $p_2\inn N\inn\veps_3,\, p_3\inn N\inn\veps_2$ are independent structures. For the other terms we use  the conservation of momentum along the brane, \ie
\beqa
(p_1+p_2+p_3)\inn V_{\mu}&=&0\labell{cons}
\eeqa
 to write $p_1^a$ in terms of $p_2^a$ and $p_3^a$.
 
The S-matrix elements of two gravitons in terms of the $RR$ structures are as following: 
  \beqa
{\cal A}_S &\sim& \eps_{a_0a_1a_2}\bigg[(p_1\inn N\inn \veps_3^S \inn N\inn \cE\inn N\inn \veps_2^S)^{a_2} \cN_1^{a_0a_1}+(p_2\inn N\inn \veps_3^S \inn N\inn \cE\inn N\inn \veps_2^S)^{a_2} \cN_2^{a_0a_1}\nonumber\\
&&\quad\quad+(p_3\inn V\inn \veps_2^S \inn N\inn \cE\inn N\inn \veps_3^S)^{a_2} \cN_3^{a_0a_1}+(p_2\inn V\inn \veps_3^S \inn N\inn \cE\inn N\inn \veps_2^S)^{a_2} \cN_4^{a_0a_1}\nonumber\\
&&\quad\quad+(p_3\inn N\inn \veps_2^S \inn N\inn \cE\inn N\inn \veps_3^S)^{a_2} \cN_5^{a_0a_1}+(p_3\inn V\inn \veps_3^S \inn N\inn \cE\inn N\inn \veps_2^S)^{a_2} \cN_6^{a_0a_1}\nonumber\\
&&\quad\quad+[(\veps_2^S \inn N\inn \cE\inn N\inn \veps_3^S)^{a_1a_2}+(2\leftrightarrow 3)]\cN_7^{a_0}+(\cE\inn N\inn \veps_3^S\inn N\inn \veps_2^S)^{a_2}\cN_8^{a_0a_1}\nonumber\\
&&\quad\quad+(\cE\inn N\inn \veps_3^S\inn V\inn \veps_2^S)^{a_2}\cN_9^{a_0a_1}+(\cE\inn N\inn \veps_3^S)^{a_2}\cN_{10}^{a_0a_1}\nonumber\\
&&\quad\quad+(\cE\inn N\inn \veps_2^S)^{a_2}\cN_{11}^{a_0a_1}\bigg]+(2\leftrightarrow 3)\labell{AlastS}
\eeqa

The subscript $S$ refers to the amplitude that involves two symmetric $NSNS$ polarization tensors. The first three lines and the first term of the fourth line of the above symmetric amplitude could capture by T-duality. One needs the T-dual Ward identity to find the other terms.  $\cN_i\,\,;i=1,\cdots,11$ ($\cM_i\,\,;i=1,\cdots,9$) are the combination of momenta, symmetric(antisymmetric) $NSNS$ polarizations and integrals functions:
\beqa
&&{\cal M}_1^{a_0a_1a_2}=p_2^{a_0}p_3^{a_1}\bigg(2(p_1\inn N\inn\veps_3^A)^{a_2}\cI_1-(p_2\inn N\inn\veps_3^A)^{a_2}\cI_2+(p_2\inn V\inn\veps_3^A)^{a_2}\cI_3-4(p_3\inn V\inn\veps_3^A)^{a_2}\cI_4\bigg)\nonumber\\
&&\quad\quad\quad\quad+p_2^{a_0} p_3\inn V\inn p_3 (\veps_3^A)^{a_1a_2}\cI_4,\nonumber\\
&&\nonumber\\
&&{\cal M}_2^{a_0a_1a_2}= p_2^{a_0}p_3^{a_1}\bigg((p_3\inn V\inn\veps_2^A)^{a_2}\cI_3-(p_3\inn N\inn\veps_2^A)^{a_2}\cI_2\bigg)\nonumber\\
&&\quad\quad\quad\quad+\frac{1}{2}p_3^{a_0}(\veps_2^A)^{a_1a_2}\bigg( p_2\inn V\inn p_3 \cI_3-p_2\inn N\inn p_3 \cI_2\bigg),\nonumber\\
&&\nonumber\\
&&{\cal N}_{11}^{a_0a_1}=p_2^{a_0}\bigg(2(p_1\inn N\inn\veps_3^S)^{a_1}\cI_1-(p_2\inn N\inn\veps_3^S)^{a_1}\cI_2+(p_2\inn V\inn\veps_3^S)^{a_0}\cI_3-p_3^{a_1}Tr(\veps_3^S\inn D)\cI_4\bigg)\nonumber\\
&&\nonumber\\
&&{\cal M}_3^{a_0a_1a_2}=2p_2^{a_0}(\veps_2^A)^{a_1a_2}\cI_1,\qquad{\cal M}_4^{a_0a_1a_2}=-p_2^{a_0}(\veps_2^A)^{a_1a_2}\cI_2,\nonumber\\
&&\nonumber\\
&&{\cal M}_5^{a_0a_1a_2}=p_2^{a_0}(\veps_2^A)^{a_1a_2}\cI_3,\qquad {\cal M}_6^{a_0a_1a_2}=-p_2^{a_0}(\veps_3^A)^{a_1a_2}\cI_2,\nonumber\\
&&\nonumber\\
&&{\cal M}_7^{a_0a_1a_2}=p_2^{a_0}(\veps_3^A)^{a_1a_2}\cI_3,\qquad{\cal N}_1^{a_0a_1}=-4p_2^{a_0}p_3^{a_1}\cI_1, \nonumber\\
&&\nonumber\\
&&{\cal M}_8^{a_0a_1}=-{\cal N}_2^{a_0a_1}={\cal N}_5^{a_0a_1}={\cal N}_9^{a_0a_1}=-2p_2^{a_0}p_3^{a_1}\cI_2\nonumber\\
&&\nonumber\\
&&{\cal M}_9^{a_0a_1}={\cal N}_3^{a_0a_1}=-{\cal N}_4^{a_0a_1}={\cal N}_8^{a_0a_1}=2p_2^{a_0}p_3^{a_1}\cI_3 \nonumber\\
&&\nonumber\\
&&{\cal N}_6^{a_0a_1}=8p_2^{a_0}p_3^{a_1}\cI_4,\qquad {\cal N}_{10}^{a_0a_1}=-2p_2^{a_0}(p_3\inn N\inn\veps_3^S)^{a_1}\cI_3\nonumber\\
&&\nonumber\\
&&{\cal N}_7^{a_0}=p_2^{a_0}\big(p_2\inn V\inn p_3\cI_3-p_2\inn N\inn p_3\cI_2+2p_3\inn V\inn p_3\cI_4\big).\qquad \nonumber\\
\eeqa
 In above sentences $\cI_2, \cI_3, \cI_4$ and $\cI_7$ are the integrals that
represent the open and closed string channels as $\cI_1$ in previous section. The explicit form of these integrals is given in \cite{Garousi:2010bm,Garousi:2011bm}. We will study these integrals and their low energy expansion in the last section. They satisfy the relation 
\beqa
-2 p_1\inn N\inn p_3 \cI_1+2 p_3\inn V\inn p_3 \cI_4+ p_2\inn N\inn p_3 \cI_2- p_2\inn V\inn p_3 \cI_3=0\labell{identity1}
\eeqa
and similar relation under the interchange of $(2\leftrightarrow 3)$. The symmetries of the integrals under the interchange of $(2\leftrightarrow 3)$ are such that $I_1$ is invariant, $\cI_2\leftrightarrow \cI_3$, and $\cI_4\leftrightarrow \cI_7$ \cite{Garousi:2011bm,Komeil:2013prd}.

 Using the above relations, one finds that the amplitudes \reef{AlastA} and \reef{AlastS} satisfy the Ward identity corresponding to
the antisymmetric and the symmetric $NSNS$ gauge transformations respectively.
 Considering the relation \reef{identity1}, it could be found that the combination of the S-matrix elements involving the last line of $\cM_1\,,\cM_2$ and the last term of amplitude \reef{A4ee} satisfies the $RR$ gauge transformations. Also the combination of the S-matrix elements involving the $\cN_7$ and the first term in amplitude \reef{A4ee} satisfies the $RR$ gauge transformations. One can check that the S-matrix elements involving all other terms of $\cM_1\,,\cM_2$ and all terms of $\cM_8\,,\cM_9,\cN_1,\cN_2\,,\cN_3,\cN_4\,,\cN_5,\cN_6,\cN_8,\cN_9$  are proportional to the RR momentum. This part of amplitude that dose not satisfy the Ward identity of the $NSNS$ gauge transformations, is in terms of $RR$ factor $(f_1)^{a_0ijk}=p_1^{a_0}\veps_1^{ijk}$. So, it could be found that the amplitudes \reef{AlastA} and \reef{AlastS}  would require the  amplitude of the RR $(p+1)$-form with lower transverse indices (in which we will study amplitude of the RR $(p+1)$-form with two transverse indices in the next section,) to become invariant under the $RR$ gauge transformations. 
 
Regardless the overall factor, these antisymmetric and symmetric amplitudes have been found by T-dual ward identity\cite{Komeil:2013prd}. T-duality relates these amplitudes to the higher RR form when the Killing coordinate carried by the NSNS polarizations\cite{Garousi:2010rn, Komeil:2013prd}.

\subsection{$n=2$ case}\labell{n2}
From the relation $n=p+1$ for this RR potential $\veps_1^{ij}$, one gets $p=1$. The amplitude corresponding  to $\veps_1^{ij}$ is nonzero only for symmetric NSNS fields and antisymmetric NSNS fields that is inherited from two previous cases.  The correlators in
$b_{4},b_{5},b_{6},b_{7},b_{8},b_{9},b_{10},b'_{13},b'_{14},b'_{15},b'_{16}$ have non-zero contribution to the amplitude \reef{A}. The trace \reef{trace} is non-zero only for $m=4$ and becomes
\beqa
T(2,1,4)&=&32\frac{4!}{2!2!}\veps_1^{ij}\eps^{a_0 a_1}A_{[ija_0 a_1]}\labell{T214}
\eeqa
calculating the sub amplitudes ${\cal A}_{4},{\cal A}_{5},{\cal A}_{6},{\cal A}_{7},{\cal A}_{8},{\cal A}_{9},{\cal A}_{10},{\cal A}'_{13},{\cal A}'_{14},{\cal A}'_{15},{\cal A}'_{16}$ in \reef{A} and combining them, one finds that the amplitude corresponding to $\veps_1^{ij}$ constructed in term of following RR  structures:
\beqa
&&Tr(\veps^A\inn N\inn \veps_1\inn N),    \quad\quad Tr(\veps^A\inn \veps^A\inn N\inn \veps_1\inn N), \quad\quad(p\inn \veps^A \inn N\inn \cE),\quad\quad \cE ,  \nonumber\\
&&(p\inn \veps^A \inn N\inn \veps_1\inn N\inn \veps^A\inn p),    \quad\quad(\cE\inn N\inn \veps^A\inn \veps^A)^{\mu},\qquad\quad(\cE\inn N\inn\veps^S)^{\mu},\nonumber\\
&&(\veps^S\inn N\inn \veps_1\inn N\inn \veps^S)^{\mu\nu},   \qquad \quad (\cE\inn N\inn \veps^S\inn \veps^S)^{\mu}, \qquad\qquad(p\inn \veps^S \inn N\inn \cE)\nonumber\\
&&(p\inn \veps^S \inn \veps^S N\inn \cE),    \quad\quad(p\inn \veps^S \inn N\inn \veps_1\inn N\inn \veps^S\inn p),    \quad\quad Tr(\veps^S\inn \veps^S\inn N\inn \veps_1\inn N),\nonumber
\eeqa
where $\mu,\nu$, as in the previous section, are the world  volume indices that contract with the volume $(p+1)$-form $\eps$. In above also, we define $\cE^{n}=(p_i\inn N\inn \veps_1)^{n},\,\, i=2,3$ that appears in contracting with NSNS polarization in the transverse direction and $\cE=p_i \inn N \inn \veps_1 \inn N \inn p_j,\,\, i,j=2,3\,\,, i\neq j$ that only appears in the last structure in the first line.

As mentioned above, the amplitude corresponding to $C^{(p+1)}$  is nonzero when both NSNS polarization tensors are either antisymmetric or symmetric. In the former case, the amplitude corresponding to $(C^{(p+1)})^{ij}$ becomes:
 \beqa
{\cal A'}_{A} &\sim & \eps_{a_0a_1}\bigg[Tr(\veps_3^A\inn N\inn \veps_1\inn N){\cal M'}_1^{a_0a_1}+\cG Tr\big((\veps_2^A V\veps_3^A)\inn N\inn\veps_1\inn N\big){\cal M'}_2^{a_0a_1}\nonumber\\
&&\quad\quad+(p_1\inn N\inn \veps_3^A \inn N \inn \cE){\cal M'}_3^{a_0a_1}+(p_1\inn N\inn \veps_2^A \inn N \inn \cE){\cal M'}_4^{a_0a_1}\nonumber\\\labell{AA}
&&\quad\quad+(p_2\inn N\inn \veps_3^A \inn N \inn \cE){\cal M'}_5^{a_0a_1}+(p_3\inn N\inn \veps_2^A \inn N \inn \cE){\cal M'}_6^{a_0a_1}\nonumber\\
&&\quad\quad+(p_2\inn V\inn \veps_3^A \inn N \inn \cE){\cal M'}_7^{a_0a_1}+(p_3\inn V\inn \veps_2^A \inn N \inn \cE){\cal M'}_8^{a_0a_1}\nonumber\\
&&\quad\quad+(p_3\inn N\inn\veps_2^A\inn N\inn \veps_1\inn N\inn \veps_3^A\inn N\inn p_2){\cal M'}_{9}^{a_0a_1}+\cE{\cal M'}_{10}^{a_0a_1}\nonumber\\
&&\quad\quad+(p_3\inn V\inn\veps_2^A\inn N\inn \veps_1\inn N\inn \veps_3^A\inn V\inn p_2){\cal M'}_{11}^{a_0a_1}\nonumber\\
&&\quad\quad+\cG(\cE \inn N\inn \veps_2^A \inn V\inn \veps_3^A)^{a_1} {\cal M'}_{12}^{a_0}\bigg]+(2\leftrightarrow 3).
\eeqa 
In the latter case the amplitude in terms of the symmetric polarization tensors is
  \beqa
{\cal A'}_S &\sim& \eps_{a_0a_1}\bigg[ (\cE \inn N\inn \veps_3^S)^{a_1}{\cal N'}_{1}^{a_0} +(\veps_2^S\inn N\inn \veps_1 \inn N\inn \veps_3^S)^{a_0a_1} \cN'_2+\cG(\cE \inn N\inn\veps_2^S\inn V\inn \veps_3^S)^{a_1}{\cal N'}_{3}^{a_0}\nonumber\\
&&\quad\quad+(p_1\inn N\inn \veps_3^S \inn N\inn \cE) {\cal N'}_{4}^{a_0a_1}+(p_2\inn N\inn \veps_3^S \inn N\inn \cE) {\cal N'}_{5}^{a_0a_1}+(p_3\inn N\inn \veps_2^S \inn N\inn \cE) {\cal N'}_{6}^{a_0a_1}\nonumber\\
&&\quad\quad+(p_2\inn V\inn \veps_3^S \inn N\inn \cE) {\cal N'}_{7}^{a_0a_1}+(p_3\inn V\inn \veps_2^S \inn N\inn \cE) {\cal N'}_{8}^{a_0a_1}+(p_3\inn V\inn \veps_3^S \inn N\inn \cE) {\cal N'}_{9}^{a_0a_1}\nonumber\\
&&\quad\quad+\cG(p_1\inn N \inn \veps_3^S\inn V\inn \veps_2^S\inn N\inn \cE){\cal N'}_{10}^{a_0a_1}+\cG(p_2\inn N \inn \veps_3^S\inn V\inn \veps_2^S\inn N\inn \cE){\cal N'}_{11}^{a_0a_1}\nonumber\\
&&\quad\quad+\cG(p_2\inn V \inn \veps_3^S\inn V\inn \veps_2^S\inn N\inn \cE){\cal N'}_{12}^{a_0a_1}+\cG(p_3\inn V \inn \veps_3^S\inn V\inn \veps_2^S\inn N\inn \cE){\cal N'}_{13}^{a_0a_1}\nonumber\\
&&\quad\quad+(p_1\inn N\inn\veps_2^S\inn N\inn \veps_1\inn N\inn \veps_3^S)^{a_1}{\cal N'}_{14}^{a_0}+(p_2\inn V\inn\veps_2^S\inn N\inn \veps_1\inn N\inn \veps_3^S)^{a_1}{\cal N'}_{15}^{a_0}\nonumber\\
&&\quad\quad+(p_3\inn N\inn\veps_2^S\inn N\inn \veps_1\inn N\inn \veps_3^S)^{a_1}{\cal N'}_{16}^{a_0}+(p_3\inn V\inn\veps_2^S\inn N\inn \veps_1\inn N\inn \veps_3^S)^{a_1}{\cal N'}_{17}^{a_0}\nonumber\\
&&\quad\quad+(p_1\inn N\inn\veps_2^S\inn N\inn \veps_1\inn N\inn \veps_3^S\inn N\inn p_1){\cal N'}_{18}^{a_0a_1}+(p_1\inn N\inn\veps_2^S\inn N\inn \veps_1\inn N\inn \veps_3^S\inn N\inn p_2){\cal N'}_{19}^{a_0a_1}\nonumber\\
&&\quad\quad+(p_1\inn N\inn\veps_2^S\inn N\inn \veps_1\inn N\inn \veps_3^S\inn V\inn p_2){\cal N'}_{20}^{a_0a_1}+(p_1\inn N\inn\veps_2^S\inn N\inn \veps_1\inn N\inn \veps_3^S\inn V\inn p_3){\cal N'}_{21}^{a_0a_1}\nonumber\\
&&\quad\quad+(p_2\inn V\inn\veps_2^S\inn N\inn \veps_1\inn N\inn \veps_3^S\inn N\inn p_2){\cal N'}_{22}^{a_0a_1}+(p_2\inn V\inn\veps_2^S\inn N\inn \veps_1\inn N\inn \veps_3^S\inn V\inn p_2){\cal N'}_{23}^{a_0a_1}\nonumber\\
&&\quad\quad+(p_2\inn V\inn\veps_2^S\inn N\inn \veps_1\inn N\inn \veps_3^S\inn V\inn p_3){\cal N'}_{24}^{a_0a_1}+(p_3\inn V\inn\veps_2^S\inn N\inn \veps_1\inn N\inn \veps_3^S\inn N\inn p_2){\cal N'}_{25}^{a_0a_1}\nonumber\\
&&\quad\quad+\cG Tr\big((\veps_2^S V\veps_3^S)\inn N\inn\veps_1\inn N\big) {\cal N'}_{26}^{a_0a_1}+\cE{\cal N'}_{27}^{a_0a_1}\bigg]+(2\leftrightarrow 3)\labell{SS}
\eeqa
where the operator $\cG$, which appears in the above amplitudes is defined as (see \cite{Komeil:2013prd})
\beqa
 {\cal G} Tr((\veps_n^A\inn V\inn\veps_m^A)\inn N\inn \veps_1\inn N )&\rightarrow &Tr((\veps_n^A\inn V\inn\veps_m^A)\inn N\inn \veps_1\inn N )-Tr((\veps_n^A\inn N\inn\veps_m^A)\inn N\inn \veps_1\inn N )\cr
 {\cal G} Tr\big((\veps_n^S \inn V\inn\veps_m^S)\inn N\inn\veps_1\inn N\big)&\rightarrow & Tr\big((\veps_n^S \inn V\inn\veps_m^S)\inn N\inn\veps_1\inn N\big)-Tr\big((\veps_n^S \inn N\inn\veps_m^S)\inn N\inn\veps_1\inn N\big)\cr
   {\cal G} (\cE \inn N\inn \veps_n^A\inn V\inn\veps_m^A )^{\mu}&\rightarrow & (\cE \inn N\inn \veps_n^A\inn V\inn\veps_m^A)^{\mu}- (\cE \inn N\inn \veps_n^A\inn N\inn\veps_m^A)^{\mu}\cr
 {\cal G} (\cE \inn N\inn\veps_n^S\inn V\inn \veps_m^S)^{\mu}&\rightarrow & (\cE \inn N\inn\veps_n^S\inn V\inn \veps_m^S)^{\mu}- (\cE \inn N\inn\veps_n^S\inn N\inn \veps_m^S)^{\mu}\cr
 {\cal G} (p\inn \veps_n^S\inn V\inn \veps_m^S\inn N\inn \cE)&\rightarrow & (p\inn \veps_n^S\inn V\inn \veps_m^S\inn N\inn \cE)- (p\inn \veps_n^S\inn N\inn \veps_m^S\inn N\inn \cE)\nonumber
\eeqa
where $n,m$ are the particle labels of NSNS polarization tensors. Acting this operator on the above structures, make them T-dual invariant. It should also be pointed
out that the T-dual effect of this operator only works for the structures appear in the third and forth line in which the Killing coordinate dose not carry with the index $\mu$. The elements of the amplitude with no RR contribution make the terms  $\cM_i\,\,;i=1,\cdots,12$ appear in the antisymmetric amplitude \reef{AA} and $\cN_i\,\,;i=1,\cdots,27$ appear in the symmetric amplitude \reef{SS}. These terms are the following: 
\beqa
{\cal M'}_1^{a_0a_1}&=&-\frac{1}{4}(p_3\inn V\inn\veps_2^A)^{a_1}\bigg(p_1\inn N\inn p_2 p_2^{a_0}\cI_3+p_1\inn N\inn p_3p_3^{a_0}\cI_2\bigg)\nonumber\\
&&-\frac{1}{4}(p_3\inn N\inn\veps_2^A)^{a_1}\bigg(p_1\inn N\inn p_3 p_2^{a_0}\cI_3+p_1\inn N\inn p_2p_3^{a_0}\cI_2\bigg)\nonumber\\
&&-p_1\inn N\inn p_3 p_3^{a_0}\bigg((p_2\inn V\inn\veps_2^A)^{a_1}\cI_7+\frac{1}{2}(p_1\inn N\inn\veps_2^A)^{a_1}\cI_1\bigg)\nonumber\\
&&-\frac{1}{8}(\veps_2^A)^{a_0a_1}\bigg(2p_1\inn N\inn p_3 p_2\inn V\inn p_2\cI_7+p_1\inn N\inn p_2 p_2\inn V\inn p_3\cI_3-p_1\inn N\inn p_2 p_2\inn N\inn p_3\cI_2\bigg)\nonumber\\
&&+\frac{1}{2}p_2^{a_0}p_3^{a_1}\bigg(p_3\inn V\inn\veps_2^A\inn V\inn p_2\cJ_1-\frac{1}{2} p_3\inn V\inn\veps_2^A\inn N\inn p_1\cI_3+p_2\inn V\inn\veps_2^A\inn N\inn p_3\cJ_2\nonumber\\
&&-p_3\inn V\inn\veps_2^A\inn N\inn p_3(\cJ+\cJ_5)+\frac{1}{2}p_3\inn N\inn\veps_2^A\inn N\inn p_1\cI_2\bigg)\nonumber\\
&&-\frac{1}{4} p_1^{a_0}(p_3\inn V\inn\veps_2^A)^{a_1}\bigg(p_2\inn V\inn p_2\cJ_1+p_3\inn V\inn p_3\cJ_4-2 p_2\inn N\inn p_3\cJ\bigg)\nonumber\\
&&-\frac{1}{2} p_1^{a_0}p_3\inn V\inn p_3\bigg((p_1\inn N\inn\veps_2^A)^{a_1}\cI_4 -(p_2\inn V\inn\veps_2^A)^{a_1}\cJ_3\bigg)\nonumber\\
&&-\frac{1}{4}p_1^{a_0}(p_3\inn N\inn\veps_2^A)^{a_1}\bigg(\cJ_{15}p_2\inn N\inn p_3+(\cJ_{16}-2\cJ)p_2\inn V\inn p_3\bigg),\nonumber\\
&&\nonumber\\
{\cal M'}_2^{a_0a_1}&=&\frac{1}{2}p_2^{a_0}p_3^{a_1}p_2\inn V\inn p_3\cJ,\qquad\qquad\nonumber\\
&&\nonumber\\
{\cal M'}_3^{a_0a_1}&=&\frac{1}{4}\bigg(p_2^{a_0}(p_3\inn V\inn\veps_2^A)^{a_1}\cI_3- 4p_2^{a_0} (p_2\inn V\inn\veps_2^A)^{a_1}\cI_4+2p_2^{a_0}(p_1\inn N\inn\veps_2^A)^{a_1}\cI_1\nonumber\\
&&+p_3^{a_0}(p_3\inn N\inn\veps_2^A)^{a_1}\cI_2+\frac{1}{2}(\veps_2^A)^{a_0a_1}\big(p_2\inn V\inn p_3 \cI_3-p_2\inn N\inn p_3 \cI_2\big)\bigg),\qquad\qquad\nonumber\\
&&\nonumber\\
{\cal M'}_4^{a_0a_1}&=&\frac{1}{8}(\veps_3^A)^{a_0a_1}\bigg(2p_2\inn V\inn p_3 \cI_4+p_2\inn V\inn p_3 \cI_3-p_2\inn N\inn p_3 \cI_2-2 p_3\inn V\inn p_3 \cI_4-4 p_1\inn N\inn p_3\cI_2\bigg)\nonumber\\
&&\qquad\quad-\frac{1}{4}\bigg(p_2^{a_0}(p_2\inn V\inn\veps_3^A)^{a_1}\cI_3+p_3^{a_0}(p_2\inn N\inn\veps_3^A)^{a_1}\cI_2-4 p_2^{a_0}(p_3\inn V\inn\veps_3^A)^{a_1}\cI_4\nonumber\\
&&\qquad\quad\quad-2p_2^{a_0}(p_1\inn N\inn\veps_3^A)^{a_1}\cI_1\bigg),\nonumber\\
&&\nonumber\\
{\cal M'}_5^{a_0a_1}&=&-\frac{1}{4}p_1^{a_0}\bigg((p_3\inn N\inn\veps_2^A)^{a_1}(\cJ_{16}-2\cJ_5)+(p_3\inn V\inn\veps_2^A)^{a_1}\cJ_{15}\bigg),\qquad\qquad\qquad\qquad\nonumber\\
&&\nonumber\\
{\cal M'}_6^{a_0a_1}&=&\frac{1}{4}p_1\inn N\inn p_3 (\veps_3^A)^{a_0a_1}\cI_3,\qquad\qquad\nonumber\\
&&\nonumber\\
{\cal M'}_7^{a_0a_1}&=&-\frac{1}{4}p_1^{a_0}\bigg((p_3\inn N\inn\veps_2^A)^{a_1}\cJ_{15}+(p_3\inn V\inn\veps_2^A)^{a_1}(\cJ_{16}-4\cJ+2\cJ_5)\bigg),\nonumber\\
&&\nonumber\\
{\cal M'}_8^{a_0a_1}&=&\frac{1}{4}p_1\inn N\inn p_3 (\veps_3^A)^{a_0a_1}\cI_2,\qquad\qquad\nonumber\\
&&\nonumber\\
{\cal M'}_{9}^{a_0a_1}&=&-{\cal M'}_{11}^{a_0a_1}=-\frac{1}{2}p_2^{a_0}p_3^{a_1}\cJ,\nonumber\\
&&\nonumber\\
 {\cal M'}_{10}^{a_0a_1}&=&\frac{1}{4}(\veps_2^A)^{a_0a_1}\bigg((p_1\inn N\inn\veps_3^A\inn N\inn p_2)\cI_2-(p_1\inn N\inn\veps_3^A\inn V\inn p_2)-2 (p_1\inn N\inn\veps_3^A\inn N\inn p_1)\cI_1\bigg)\nonumber\\
&&-\frac{1}{4}p_1^{a_0}\bigg(\cG(p_2\inn N\inn \veps_3^A\inn V\inn \veps_2^A)^{a_1}\cJ_{15} +\cG(p_2\inn V\inn \veps_3^A\inn N\inn \veps_2^A)^{a_1}\cJ_{15}\nonumber\\
&&-\cG(p_2\inn V\inn \veps_3^A\inn V\inn \veps_2^A)^{a_1}(4\cJ-\cJ_{16}-2\cJ_{5})+\cG(p_2\inn N\inn \veps_3^A\inn N\inn \veps_2^A)^{a_1}(\cJ_{16}-2\cJ_{5})\bigg)\nonumber\\
&&-2\bigg(p_1^{a_0}\cG(p_1\inn N\inn \veps_3^A\inn N\inn \veps_2^A)^{a_1}\cI_{2}+p_2^{a_0}\cG(p_1\inn N\inn \veps_3^A\inn V\inn \veps_2^A)^{a_1}\cI_{3}\bigg)\nonumber\\
&&-\frac{1}{2}p_2^{a_0}p_3^{a_1}\bigg((2\cJ-\cJ_{16})(\Tr(\veps_2^A\inn N\inn \veps_3^A\inn N)+\Tr(\veps_2^A\inn V\inn \veps_3^A\inn V))\nonumber\\
&&-2\cJ_{15}\Tr(\veps_2^A\inn V\inn \veps_3^A\inn N)\bigg),\nonumber\\
&&\nonumber\\
{\cal M'}_{12}^{a_0}&=&\frac{1}{2}\bigg(p_2^{a_0}p_1\inn N\inn p_3\cI_2+p_1^{a_0}p_3\inn V\inn p_3\cJ_{12}\bigg),\nonumber\\
&&\nonumber\\
{\cal N'}_{1}^{a_0}&=&\frac{1}{2}p_2^{a_0}\bigg( p_1\inn N\inn\veps_2^S \inn V\inn p_3\cI_3+4 p_1\inn N\inn\veps_2^S \inn V\inn p_2\cI_4-2 p_1\inn N\inn\veps_2^S \inn N\inn p_1\cI_1\bigg)\nonumber\\
&&- p_3^{a_0} p_1\inn N\inn\veps_2^S \inn N\inn p_3\cI_2-\frac{1}{4}(p_1\inn N\inn\veps_2^S)^{a_0}\bigg(2p_3\inn V\inn p_3\cI_4+p_2\inn V\inn p_3\cI_3-p_2\inn N\inn p_3\cI_2\bigg)\nonumber\\
&&-\frac{1}{2}(p_3\inn N\inn\veps_2^S)^{a_0} p_1\inn N\inn p_3\cI_3+\frac{1}{2}(p_3\inn V\inn\veps_2^S)^{a_0} p_1\inn N\inn p_3\cI_2+p_2^{a_0}Tr(\veps_2^A\inn V)p_1\inn N\inn p_3\cI_7\nonumber\\
&&-\frac{1}{4}p_1^{a_0}\bigg(2p_2\inn V\inn\veps_3^S\inn N\inn p_2\cJ_{15}-2(2\cI_{4}p_1\inn N\inn p_3-\cJ_4p_2\inn N\inn p_3+\cJ_{12}p_2\inn V\inn p_3)\Tr[\veps_3^S\inn V]\nonumber\\
&&+p_2\inn V\inn\veps_3^S\inn V\inn p_2(\cJ_{16}-4\cJ+2\cJ_5)+4p_3\inn V\inn\veps_3^S\inn V\inn p_2\cJ_{12}-2 p_2\inn V\inn\veps_3^S\inn N\inn p_1\cI_3\nonumber\\
&&+p_2\inn N\inn\veps_3^S\inn N\inn p_2(\cJ_{16}-2\cJ_5)-4\cJ_{4} p_3\inn V\inn\veps_3^S\inn N\inn p_2+4\cI_2p_2\inn N\inn\veps_3^S\inn N\inn p_1\bigg),\nonumber\\
&&\nonumber\\
{\cal N'}_2&=&-\frac{1}{4}\bigg(2p_1\inn N\inn p_2 p_3\inn V\inn p_3\cI_4+p_1\inn N\inn p_3 p_2\inn V\inn p_3\cI_2-p_1\inn N\inn p_3 p_2\inn N\inn p_3\cI_3\bigg),\nonumber\\
&&\nonumber\\
{\cal N'}_3^{a_0}&=&\frac{1}{2}\bigg(p_1^{a_0}p_3\inn V\inn p_3\cJ_{12}+p_2^{a_0}p_1\inn N\inn p_3\cI_2\bigg),\nonumber\\
&&\nonumber\\
{\cal N'}_{4}^{a_0a_1}&=&p_2^{a_0}(p_1\inn N\inn\veps_2^S)^{a_1}\cI_1,\nonumber\\
&&\nonumber\\
 {\cal N'}_{5}^{a_0a_1}&=&-\frac{1}{2}p_3^{a_0}(p_1\inn N\inn\veps_2^S)^{a_1}\cI_2,\nonumber\\
&&\nonumber\\
 {\cal N'}_{6}^{a_0a_1}&=&-\frac{1}{4}p_1^{a_0}\bigg(2p_3^{a_1}\Tr[\veps_3^S\inn V]\cJ_{4}-2(p_1\inn N\inn\veps_3^S)^{a_1} \cI_{2}-(p_2\inn N\inn\veps_3^S)^{a_1}(\cJ_{16}-2\cJ_5)\nonumber\\
&&-(p_2\inn V\inn\veps_3^S)^{a_1}\cJ_{15}\bigg),\nonumber\\
&&\nonumber\\
 {\cal N'}_{7}^{a_0a_1}&=&-\frac{1}{2}p_2^{a_0}(p_1\inn N\inn\veps_2^S)^{a_1}\cI_3,\nonumber\\
&&\nonumber\\
 {\cal N'}_{8}^{a_0a_1}&=&\frac{1}{4}p_1^{a_0}\bigg(2p_3^{a_1} \Tr[\veps_3^S\inn V]\cJ_{12}+(p_2\inn N\inn\veps_3^S)^{a_1}\cJ_{15}+(p_2\inn V\inn\veps_3^S)^{a_1}(\cJ_{16}-4\cJ+2\cJ_5)\bigg),\nonumber\\
&&-\frac{1}{2}p_2^{a_0}(p_1\inn N\inn\veps_3^S)^{a_1}\cI_3,\nonumber\\
&&\nonumber\\
{\cal N'}_{9}^{a_0a_1}&=&2p_2^{a_0}(p_1\inn N\inn\veps_2^S)^{a_1}\cI_4,\nonumber\\
&&\nonumber\\
{\cal N'}_{10}^{a_0a_1}&=&-\frac{1}{2} p_2^{a_0}p_3^{a_1}(\cI_{3}+\cI_2),\nonumber\\
&&\nonumber\\
{\cal N'}_{11}^{a_0a_1}&=&\frac{1}{4} p_2^{a_0}p_3^{a_1}\bigg(\cJ_{15}-\cJ_{16}+2\cJ_5\bigg),\nonumber\\
&&\nonumber\\
{\cal N'}_{12}^{a_0a_1}&=&\frac{1}{4}p_2^{a_0}p_3^{a_1}\bigg(\cJ_{16}-4\cJ+2\cJ_5 -\cJ_{15}\bigg),\nonumber\\
&&\nonumber\\
{\cal N'}_{13}^{a_0a_1}&=&p_2^{a_0}p_3^{a_1}(\cJ_{12}+\cJ_4),\nonumber\\
&&\nonumber\\
{\cal N'}_{14}^{a_0}&=&p_1^{a_0}p_3\inn V\inn p_3\cI_4-\frac{1}{2} p_3^{a_0}p_1\inn N\inn p_3 \cI_1,\nonumber\\
&&\nonumber\\
{\cal N'}_{15}^{a_0}&=&2\bigg(p_1^{a_0}p_3\inn V\inn p_3\cJ_3+p_3^{a_0}p_1\inn N\inn p_3 \cI_7\bigg),\nonumber\\
&&\nonumber\\
{\cal N'}_{16}^{a_0}&=&-\frac{1}{2}\bigg(p_1^{a_0}(\cJ_{15}p_2\inn N\inn p_3+(\cJ_{16}-2\cJ)p_2\inn V\inn p_3)+p_3^{a_0}p_1\inn N\inn p_2 \cI_2+p_2^{a_0}p_1\inn N\inn p_3\cI_3\bigg),\nonumber\\
&&\nonumber\\
{\cal N'}_{17}^{a_0}&=&\frac{1}{2}\bigg(p_1^{a_0}(p_2\inn V\inn p_2\cJ_1+p_3\inn V\inn p_3\cJ_4-2p_2\inn N\inn p_3\cJ)+p_1\inn N\inn p_2 p_2^{a_0}\cI_3+p_1\inn N\inn p_3p_3^{a_0}\cI_2\bigg),\nonumber\\
&&\nonumber\\
{\cal N'}_{18}^{a_0a_1}&=&\frac{1}{2}p_2^{a_0}p_3^{a_1}\cI_1,\qquad\qquad
{\cal N'}_{19}^{a_0a_1}=-\frac{1}{2}p_2^{a_0}p_3^{a_1}\cI_2,\qquad\qquad
{\cal N'}_{20}^{a_0a_1}=\frac{1}{2}p_2^{a_0}p_3^{a_1}\cI_3,\nonumber\\
&&\nonumber\\
{\cal N'}_{21}^{a_0a_1}&=&-2p_2^{a_0}p_3^{a_1}\cI_4,\qquad\qquad
{\cal N'}_{22}^{a_0a_1}=p_2^{a_0}p_3^{a_1}\cJ_2,\qquad\qquad
{\cal N'}_{23}^{a_0a_1}=-p_2^{a_0}p_3^{a_1}\cJ_1,\nonumber\\
&&\nonumber\\
{\cal N'}_{24}^{a_0a_1}&=&2 p_2^{a_0}p_3^{a_1}\cJ_3,\qquad\qquad
{\cal N'}_{25}^{a_0a_1}=p_2^{a_0}p_3^{a_1}\cJ_5,\qquad\qquad
{\cal N'}_{26}^{a_0a_1}=-\frac{1}{2}p_2^{a_0}p_3^{a_1}p_2\inn N\inn p_3\cJ,\nonumber\\
&&\nonumber\\
 {\cal N'}_{27}^{a_0a_1}&=&-\frac{1}{2}\Tr(\veps_3^S\inn V)\bigg(-2p_3^{a_0}\cI_4(p_1\inn N\inn \veps_2^S)^{a_1}\nonumber\\
&&+p_1^{a_0}(\cJ_4(p_3\inn V\inn \veps_2^S)^{a_1}-\cJ_{12} (p_3\inn N\inn \veps_2^S)^{a_1}+\cJ_3 p_3^{a_1}\Tr(\veps_2^S\inn V))\bigg)\nonumber\\
&&-\frac{1}{2}(p_1\inn N\inn \veps_2^S)^{a_0}\bigg( (p_2\inn N\inn\veps_3^S)^{a_1}\cI_2 +(p_2\inn V\inn\veps_3^S)^{a_1}\cI_3-2 (p_1\inn N\inn\veps_3^S)^{a_1}\cI_1\bigg)\nonumber\\
&&-\frac{1}{2}p_2^{a_0}p_3^{a_1}\bigg(-2\Tr(\veps_2^S\inn V\inn \veps_3^S\inn N)(2\cJ-\cJ_{16})+\Tr(\veps_2^S\inn N\inn \veps_3^S\inn N)\cJ_{15}\nonumber\\
&&+\Tr(\veps_2^S\inn V\inn \veps_3^S\inn V)\cJ_{15}\bigg)+2p_3^{a_0}\bigg(\cG(p_1\inn N\inn \veps_2^S\inn V\inn\veps_3^S)^{a_1}\cI_3\bigg).\nonumber
\eeqa
In  the above  sentences $\cJ$ and $\cJ_i$'s are some other integrals as $\cI_i$'s in the previous sections in which the explicit forms are given in \cite{Garousi:2010bm, Komeil:2013prd}. The symmetries of the integrals under the interchange of $(2\leftrightarrow 3)$ are such that $\cJ,\cJ_3,\cJ_{15},\cJ_{16}$ are invariant,  $\cJ_{1}\leftrightarrow\cJ_4 $,  $\cJ_{2}\leftrightarrow\cJ_{12}$, and $\cJ_{5}\leftrightarrow-\cJ_5 $. The antisymmetric amplitude \reef{AA} and the symmetric amplitude \reef{SS} satisfy the Ward identity corresponding to the B-field and graviton gauge transformations respectively, provided the integrals satisfy the relations
\beqa
-2 \cI_3 p_1\inn N\inn p_3+\cJ_{15}  p_2\inn N\inn p_3+2 \cJ_{12} p_3\inn V\inn p_3 +(-4 \cJ+\cJ_{16} +2 \cJ_{5}) p_2\inn V\inn p_3&=&0\nonumber\\
2 \cI_{2} p_1\inn N\inn p_3-2 \cJ_{4} p_3\inn V\inn p_3+\cJ_{15}  p_2\inn V\inn p_3 +(\cJ_{16} -2 \cJ_{5}) p_2\inn N\inn p_3&=&0\nonumber\\
-2 \cI_{7} p_1\inn N\inn p_3+\cJ_{2} p_2\inn N\inn p_3+2 \cJ_{3} p_3\inn V\inn p_3-\cJ_{1} p_2\inn V\inn p_3&=&0\labell{iden3}
\eeqa
and similar relations under the interchange of $(2\leftrightarrow 3)$. We check the above identities by performing the integrals explicitly.



The amplitudes \reef{AA} and \reef{SS} are not invariant under the RR ward identity. Combining the part of the amplitudes \reef{AA} and \reef{SS} involving the first three lines in $\cN'_1$, the last terms in $\cN'_{3},\cN'_{8},\cN'_{14},\cN'_{15}$, the last two terms in $\cN'_{16},\cN'_{17}$, the terms in $\cN'_{4},\cN'_{5},\cN'_{7},\cN'_{9}$, the first term in $\cN'_{27}$, the first four lines in $\cM'_1$, the terms in $\cM'_{3},\cM'_{4},\cM'_{6},\cM'_{8}$ and the first term in $\cM'_{12}$ with the amplitude corresponding to $C^{(p+1)}$ with three transverse indices, which has been found in the previous section, satisfy the RR Ward identity.  The other terms in all $\cN_i$'s and $\cM_i$'s proportional to RR momentum $p_1$. The amplitude corresponding to $C^{(p+1)}$ with three transverse indices has also some terms that satisfy the RR Ward identity after combining with the amplitude corresponding to $C^{(p+1)}$ with four transverse indices which has been found in the section \reef{n4}. However the RR invariant amplitude dose not satisfy the Ward identity associated with the NSNS gauge transformations anymore. The asymmetry of the amplitudes corresponding to $C^{(p+1)}$ with four, three and two transverse indices under the RR gauge transformations indicates that one has to include the amplitude corresponding to $C^{(p+1)}$ with one transverse index in which we are not interested.

The polarization tensors in all terms except the terms including the operator $\cG$, are contracted with the momentum or with the world volume form. On the other hand, as we mentioned before, the operator $\cG$ generate the T-dual invariant structure. So the amplitudes \reef{AA} and \reef{SS} invariant under linear T-duality when the Killing coordinate is an index of the RR potential. T-duality relates the amplitudes to the higher RR form when the Killing coordinate carries by the NSNS polarizations.

 Applying the linear T-duality on the amplitude of $C^{(p-1)}$ that has one transverse index, in which has been found using T-dual Ward identity in \cite{Komeil:2013prd} and explicitly in \cite{Komeil:2015}, when the Killing coordinate is carried by the NSNS polarization tensors, regardless the overall factor, one can find the amplitude \reef{AA} and \reef{SS} except the elements constructed with the last four lines in $\cM'_{10}$, the second line, the fourth line and the first term in the last line in $\cN'_{27}$. It could be found that T-duality can only predict the structures contribute to these above-mentioned terms  in $\cM'_{10}$ and $\cN'_{27}$ and can not capture them with the right integral functions.

\section{Four derivative couplings}
The amplitudes corresponding to RR $(p+1)$-form potential $C^{(p+1)}$ with four, three and two transverse indices that we have found in this paper are invariant under the Ward identity associated with the B-field and graviton gauge transformations. Hence, it could be rewrite these amplitudes in terms of the field strength of the NSNS states. One cane rewrite the amplitude \reef{A4ee} in terms of $H$ and $R$ as
\beqa
\cal {A}&\sim& T_3\eps_{a_0\cdots a_3}\bigg((\cR_2\inn N\inn N\inn\veps_1\inn N\inn N\inn\cR_3)^{a_0a_1a_2a_3}+\frac{1}{6}p_2^{a_0}\Tr(\cE\inn N\inn N\inn N\inn H_2\inn N\inn N\inn N) H_3^{a_1a_2a_3}\bigg)\nonumber\\\labell{A4HR}
&&\times\cI_1\delta(p_1^a+p_2^a+p_3^a)+(2\leftrightarrow 3)
\eeqa

where $\cE^{nmr}=(p_3\inn N\inn \veps_1)^{nmr}$. Our notation for two(three) consecutive projection operators $N\inn N$($N\inn N\inn N$) in the above amplitude and in the following  in this paper is that the two (three) transverse legs of involving tensors contracted with each other \ie
$(\veps_1)_{ijkl}\cR_2^{a_0a_1ij}\cR_3^{a_2a_3kl}=(\cR_2\inn\, N\inn\, N\inn\,\veps_1\inn\, N\inn\, N\inn\,\cR_3)^{a_0a_1a_2a_3}$ and $p_3^{l} (\veps_1)_{ijkl}H_2^{ijk}=\Tr(\cE\inn\, N\inn\, N\inn\, N\inn\, H_2\inn\, N\inn\,N\inn\,N)$.

The next amplitude that we want to write in terms of NSNS fields strength is the $(\veps_1^{(p+1)})^{ijk}$ amplitude. This amplitude has two parts, the antisymmetric amplitude  \reef{AlastA} and symmetric amplitude \reef{AlastS}. Using the identity \reef{identity1}, one can rewrite the former amplitude in terms of $H_2$ and $H_3$ as
\beqa
{\cA}^A&\sim &T_2\eps_{a_0a_1a_2}\bigg[\bigg(2p_1\inn N\inn H_3\inn N\inn N\inn \cE\, \cI_1- p_2\inn N\inn H_3\inn N\inn N\inn \cE\, \cI_2+p_2\inn V\inn H_3\inn N\inn N\inn \cE\, \cI_3\bigg)H_2^{a_2a_1a_0}\nonumber\\
&&+p_2^{a_2}\bigg(2(p_1\inn N\inn H_3)^{a_1a_0}\cI_1- (p_2\inn N\inn H_3)^{a_1a_0}\cI_2+(p_2\inn V\inn H_3)^{a_1a_0}\cI_3-4(p_3\inn V\inn H_3)^{a_1a_0}\cI_4\nonumber\\\labell{AAHH}
&&-\frac{1}{3}H_3^{a_2a_1a_0}\big((p_2\inn N\inn p_3)\cI_2- (p_2\inn V\inn p_3)\cI_3\big)\bigg)\Tr(\veps_1\inn N\inn N\inn N\inn H_2\inn N\inn N\inn N)\nonumber\\
&& + 3p_2^{a_2}\bigg((\cE\inn N\inn N\inn H_2\inn V\inn H_3)^{a_1a_0}\cI_2-(\cE\inn N\inn N\inn H_2\inn N\inn H_3)^{a_1a_0}\cI_3\bigg) \bigg]+(2\leftrightarrow 3)
\eeqa
where $\cE^{nm}=(p_2\inn N\inn \veps_1)^{nm}$ in the first line and $\cE^{nm}=(p_3\inn N\inn \veps_1)^{nm}$ in the last line. As we know, the metric contribute to curvature tensor and covariant derivative in the effective action. The metric in the effective action also appears in contracting the indices. Because of these points, one dose not expect that all the contact terms of the symmetric amplitude \reef{AlastS} to be rewritten as $R_2R_3$. So we left with this amplitude in terms of it's symmetric NSNS polarizations.  

Before going to the next amplitude that we found in the section \reef{n2}, Let us analyze the above amplitudes at low energy to find the coupling of one RR $(p+1)$-form and two NSNS states. So we need the $\alpha'$ expansion of the integral functions that appear in the amplitudes. These low energy expansion contain massless poles as well as contact terms that have been found in \cite{Becker:2011ad}.

 Some contact terms, that may come from the subtraction of field theory massless poles from the string theory amplitude, add to the contact terms of the string theory amplitude. For antisymmetric amplitudes which one can write in terms of the antisymmetric NSNS fields, it may be expected that this additional contact terms to be avoided when one writes
the string theory amplitude,
in terms of $H$\cite{Garousi:2010bm,Garousi:2011bm}. 

Using this point and considering the contact terms in the low energy expansion of the integrals functions, one can find couplings at order $\alpha'^2$. It has been shown in \cite{Becker:2011ad} that the low energy expansion of the integral functions $\cI_1\,,\cI_2$ and $\cI_4$ have no contact term contribution. So the low energy limit of the amplitudes \reef{A4HR} and \reef{AAHH} contain any couplings. 

Now consider the last case of the amplitudes in this paper where the RR $p+1$-form carries two transverse indices in which we found it as antisymmetric amplitude \reef{AA} and symmetric amplitude \reef{SS}. In the latter amplitude also one cannot rewrite it in terms of curvatures tensors. In the former amplitude, it can be rewritten in terms of antisymmetric NSNS fields strength. The integral functions $\cJ_3$ and $\cJ_5$ also have any contact term in their low energy expansions \cite{Becker:2011ad}. Hence, without considering the terms containing the integral functions $\cJ_3$ and $\cJ_5$, one can rewrite the amplitude \reef{AA} in terms of $H_2$ and $H_3$ . The result is  
\beqa
 {\cA}^A&\sim &T_1\eps_{a_0a_1}\bigg[ (-4\cJ+\cJ_{16}) p_3^{a_0} (H_2\inn V\inn V\inn H_3\inn N\inn \cE)^{a_1}+2 \cJ_{15} p_3^{a_0} (\cE \inn N\inn H_3\inn V\inn N\inn H_2)^{a_1}\nonumber\\
&&+ (-4\cJ+\cJ_{16}) p_3^{{a_0}} (\cE \inn N\inn H_3\inn N\inn N\inn H_2)^{a_1}+4 {\cJ_2} p_3^{a_0}(p_2\inn V\inn{H_2}\inn N\inn {H_3}\inn N\inn N\inn\veps_1)^{{a_1}}\nonumber\\
&&-4{\cJ_1} p_3^{a_0}(p_2\inn V\inn{H_2}\inn V\inn{H_3}\inn N\inn N\inn\veps_1)^{{a_1}}+\cJ_{15} p_3^{a_0} (p_3\inn N\inn{H_2}\inn N\inn{H_3}\inn N\inn N\inn\veps_1)^{{a_1}}\nonumber\\
&&-8\cJ p_2^{{a_0}} p_3\inn N\inn H_2\inn{\cal{N}} \inn N \inn H_3^{a_1}\inn N \inn ^{}\veps_1\inn{\cal{N}}+\cJ_{16}  p_2\inn N\inn H_3\inn{\cal{N}} \inn N \inn H_2^{a_0a_1}\cE\inn{\cal{N}}\nonumber\\
&&+ \cJ_{16} p_3^{a_0} (\veps_1\inn N\inn N\inn H_3\inn V\inn H_2\inn N\inn p_3)^{a_1}+ \cJ_{15} p_2\inn N\inn H_3 \inn{\cal{N}}\inn V \inn H_2^{a_0a_1}\cE\inn{\cal{N}}\nonumber\\
&&+ (-4 \cJ+\cJ_{16}) p_3^{a_0}(\veps_1\inn N\inn N\inn H_3\inn N\inn H_2\inn V\inn p_3)^{a_1}+\cJ_{15} p_2\inn V\inn H_3\inn{\cal{N}} \inn N \inn H_2^{a_0a_1} \nonumber\\
&&+\cJ_{15} p_3^{a_0}(\veps_1\inn N\inn N\inn H_3\inn V\inn H_2\inn V\inn p_3)^{a_1}+8\cJ p_2^{{a_0}} p_3\inn V\inn H_2\inn{\cal{N}} \inn V \inn H_3^{a_1}\inn N\inn\veps_1\inn{\cal{N}}\nonumber\\
&&+\cJ_{12} p_3\inn V\inn p_3(\veps_1 \inn N\inn N\inn {H_2}\inn N\inn{H_3})^{{a_0}{a_1}}+ (-4\cJ+\cJ_{16}) p_2\inn V\inn H_3\inn{\cal{N}} \inn V \inn H_2^{a_0a_1}\nonumber\\
&& -{\cJ_4} p_3\inn V\inn p_3  (\veps_1 \inn N\inn N\inn{H_2}\inn V\inn{H_3})^{{a_0}{a_1}}
-\frac{1}{16}  (p_3\inn N\inn{H_2}\inn N\inn N\inn \veps_1)\bigg((p_2\inn N\inn{H_3})^{{a_0}{a_1}}\cJ_{15}\nonumber\\
&&-(p_2\inn V\inn{H_3})^{{a_0}{a_1}} (4 \cJ-\cJ_{16})    +4 (p_3\inn V\inn{H_3})^{{a_0}{a_1}} \cJ_{12} \bigg)-\frac{1}{16}  (p_3\inn V\inn{H_2}\inn N\inn N\inn \veps_1) \nonumber\\
&&\bigg((p_2\inn V\inn{H_3})^{{a_0}{a_1}}\cJ_{15} - (p_2\inn N\inn{H_3})^{{a_0}{a_1}}\cJ_{16}-4 (p_3\inn V\inn{H_3})^{{a_0}{a_1}} \cJ_{4}\bigg)\bigg]+(2\leftrightarrow 3)
\eeqa  
where $\cE^{n}=(p_2\inn N\inn \veps_1)^{n}$. The notation for the appearance of two different consecutive projection operators in the structure $(\cE \inn N\inn H_3\inn V\inn N\inn H_2)^{a_1}$, for example, indicate that one transverse leg and one world volume leg of $H_2$ contract with $H_3$ and another leg carries the external index $a_1$ in which contract with the volume $p+1$-form tensor. (For another example, in the first structure in the sixth line, three transverse legs of $H_2$ contract with $p_3\,,H_3$ and $\veps_1$). $N$ and $\cN$ were defined to indicate different transvers contractions.
\newpage

From the contact term contributions in the low energy expansion of the integrals in the above amplitude, one can find the contact amplitude. The Lagrangian corresponding to this contact amplitude becomes
\beqa 
\cL &\sim &\frac{ (\pi ^2\alpha')^2T_1}{6}\eps_{a_0a_1} (C^{(p+1)})_{ij}\nonumber\\
&&\times\bigg[\frac{1}{2} H^{nij,m}H^{ma_0a_1,n} -\frac{1}{2} H^{nij,a}H^{aa_0a_1,n} - H^{nij}H^{aa_0a_1,an}\nonumber\\
&&+\frac{1}{2} H^{aij,b}H^{ba_0a_1,a}+\frac{1}{2} H^{aij,n}H^{na_0a_1,a}- H^{aij}H^{ba_0a_1,ab}\nonumber\\
&&  + H^{aba_1,i} H^{abj,a_0}-2 H^{anj,a_0}H^{ana_1,i}+H^{nmj,a_0}H^{nma_1,i}\nonumber\\
&& + 2 H^{ana_1,a} H^{nij,a_0}+ 2 H^{aba_1,a} H^{bij,a_0}- H^{nma_1} H^{mij,na_0}\nonumber\\
&& + 4 H^{nmi,a_0}H^{mja_1,n}-H^{nmj}H^{ma_0a_1,in}-H^{naa_1}H^{aij,na_0}\nonumber\\
&& -H^{naj}H^{aa_0a_1,ni}+H^{ana_1}H^{nij,aa_0}-H^{anj}H^{na_0a_1,ai}\nonumber\\
&& - H^{aba_1}H^{bij,aa_0}-4H^{abi,a_0}H^{bja_1,a}+H^{abj}H^{a_0a_1b,ai}\nonumber\\
&&  +\frac{1}{2}H^{ij\mu}H^{a_0a_1\mu,aa}\bigg]+(2\leftrightarrow 3)
\eeqa
The first term in the fifth line and the second term in the seventh line are exactly the first terms in the first and second lines of the coupling \reef{new25} respectively. All other terms are new couplings which should be added to the D-brane action at order $\alpha'^2$.

We have found the couplings $C^{(p+1)}\wedge H\wedge H$ by using the explicit calculation of relevant scattering amplitude. Comparing the string low energy amplitude with the Field theory amplitude, one can find all nonzero coupling for two graviton in this case. On the other hand, when one of the indices of the B-field polarization tensors which is contracted with the volume form is the Killing coordinate, the above couplings are not invariant under the linear T-duality. In that case, one should add new couplings involving higher RR potential to make a complete T-dual multiplet. It would be interesting to find this multiplet.

{\bf Acknowledgments}:  We would like to thank M.R.Garousi for very valuable discussions. This work is supported by University of Guilan.

\end{document}